\documentclass[preprint2]{aastex}

\usepackage{graphicx}
\usepackage{txfonts}
\usepackage{array}
\usepackage{amssymb}
\usepackage{natbib}
\usepackage{lscape}
\usepackage{rotating}
\usepackage{longtable}
\newcommand{\hd}{HD\,191612}
\newcommand{\ch}{{\it Chandra}}
\newcommand{\xmm}{{\sc{XMM}}\emph{-Newton}}
\newcommand{\kms}{km\,s$^{-1}$}

\begin{document}

\title{ \ch\ View of Magnetically Confined Wind in \hd: Theory versus Observations}

\author{Ya\"el Naz\'e\altaffilmark{1,2}}
\affil{GAPHE - STAR - Institut d'Astrophysique et de G\'eophysique (B5C), Universit\'e de Li\`ege, All\'ee du 6 Ao\^ut 19c, 4000-Li\`ege, Belgium}
\author{Asif ud-Doula}
\affil{Penn State Worthington Scranton, Dunmore, PA 18512, USA}
\and
\author{Svetozar A. Zhekov}
\affil{Institute of Astronomy and National Astronomical Observatory, 72 Tsarigradsko Chaussee Blvd., Sofia 1784, Bulgaria}
\altaffiltext{1}{FNRS Research Associate}
\altaffiltext{2}{naze@astro.ulg.ac.be}

\begin{abstract}
High-resolution spectra of the magnetic star \hd\ were acquired using the \ch\ X-ray observatory at both maximum and minimum emission phases. We confirm the flux and hardness variations previously reported with \xmm, demonstrating the great repeatability of the behavior of \hd\ over a decade. The line profiles appear typical for magnetic massive stars: no significant line shift, relatively narrow lines for high-Z elements, and formation radius at about 2$R_*$. Line ratios confirm the softening of the X-ray spectrum at the minimum emission phase. Shift or width variations appear of limited amplitude at most (slightly lower velocity and slightly increased broadening at minimum emission phase, but within 1--2$\sigma$ of values at maximum). In addition, a fully self-consistent 3D magnetohydrodynamic (MHD) simulation of the confined wind in \hd\ was performed. The simulation results were {\it directly} fitted to the  data leading to a remarkable agreement overall between them.
\end{abstract}


\section{Introduction}

Classified as Of?p nearly half a century ago \citep{wal73}, \hd\ regained interest only a decade ago when large variations of its line profiles were identified \citep{wal03}. As in HD\,108, another Of?p star \citep{naz01}, strong narrow emissions, especially in H,He\,{\sc i} lines, practically disappear at certain times. A photometric period of $\sim$537d was then identified for \hd, thanks to Hipparcos photometry \citep{koe02,naz04}, and it was readily shown to be consistent with the spectroscopic changes \citep{wal04,how07}. Moreover, variations of the X-ray flux \citep{naz07,naz10} and UV line profiles \citep{mar13} were detected, and found to occur in phase with those in the visible range. Finally, \hd\ became the second O-star with a detected (strong) magnetic field \citep{don06,wad11}. Currently, there are about a dozen known O stars with detectable global magnetic field \citep{fos15,wad16}.

The field detection appeared as a key to understand the star's peculiarities. Indeed, such a strong magnetic field is able to channel the stellar winds from opposite hemispheres towards the equatorial regions, forming a disk-like feature. This slow-moving, dense material generates narrow emissions in the visible range (notably in H,He\,{\sc i} lines). The variations detected at optical wavelengths could be closely reproduced with magnetohydrodynamic (MHD) models simply by changing the angle-of-view on the confined winds \citep{sun12}. Indeed, when the rotation and magnetic axes are not aligned, our view towards these magnetically-confined winds changes with time. This can also explain the behavior in UV as confined winds seen edge-on are able to produce the larger absorption at low velocities seen in the UV profiles \citep{mar13}. Finally, the collision between the wind flows is able to produce multi-million degree plasma, generating X-ray emission \citep{bab97}. Depending on geometry, some occultation may occur as the stellar body comes into the line-of-sight towards the confined winds at some (rotational) phases.

While improvements in our understanding of confined winds have been tremendous in the last decade, several aspects remain to be explained. To further gain insight on the hottest plasma in magnetospheres, high-resolution X-ray spectra with different angles-of-view on the magnetosphere are needed. Few strongly magnetic O-stars can be studied this way, however. Most objects (e.g. NGC1624-2 \citealt{pet15}, CPD --28$^{\circ}$2561 \citealt{naz15}, HD\,57682 \citealt{naz14}, Tr16-22 \citealt{naz14tr16}) are much too faint for such an endeavor, while others have more practical problems - e.g. the long period (about 55\,yrs, \citealt{naz06}) of HD\,108 prohibits a study of its variability over the lifetime of X-ray satellite missions. Currently, a high-resolution spectral analysis of confined winds is thus possible only for three stars: $\theta^1$\,Ori\,C, \hd, and HD\,148937. 

The latter object, discussed in \citet{naz12,naz14}, has a constant X-ray emission, linked to a quasi unchanged view of its magnetosphere (always seen near pole-on) which limits the available information. High-resolution spectral analysis of the O star $\theta^1$\,Ori\,C  is also available \citep[e.g.][]{sch03,gag05}. In particular, \citet{gag05} showed  that `magnetically confined wind shock' (MCWS) paradigm \citep{bab97} was clearly at work even in an O star. However, their analysis was based on, although fully self-consistent, 2D MHD simulations which naturally impose an artificial azimuthal symmetry. Furthermore, they compared their numerical models to the observational data  only {\it indirectly}: for example, temperatures and line widths derived from XSPEC fits were confronted to values independently estimated from simulation outputs -- the numerical model itself was never directly fitted to the observational data to judge its adequacy.

Work here reflects further improvements in several aspects. A \ch\ monitoring of \hd\ at high-resolution allows us to see the hottest magnetospheric component under different angles, leading to precise observational constraints of its properties. We present the first fully self-consistent 3D MHD model of \hd.  Relying on the dynamical output of this numerical model, we use a dedicated XSPEC model to make a {\it direct} comparison between the theory and observations.
 
In the next section, we present the observations and their reduction. This is followed by a discussion of our 3D MHD model. We then present the results, including the direct comparison between the observations and our models, in \S 4, and we summarize our results in \S 5.

\section{Observations and data reduction}

High-resolution spectroscopy of \hd\ was acquired with \ch -HETG at two key phases, the maximum and minimum emission phases. These phases correspond to specific angles-of-view onto the confined winds. Indeed, \citet{wad11} derived $\beta+i=95\pm10^{\circ}$ (with $i$ the inclination angle and $\beta$ the obliquity of the magnetic axis relative to the rotation axis), while \citet{sun12} showed that $\beta=i$ yielded the best fit to the variations in the strength of H$\alpha$ emission component. Thus, the maximum emission phase corresponds to a pole-on view of \hd, with confined winds seen face-on, while minimum emission corresponds to an equatorial view, with confined winds disk-like structure seen edge-on. 

The maximum was covered by four exposures in May-July 2015 totaling 142\,ks, while the observation at minimum was split over 6 exposures in early 2016 totaling 196\,ks (Table \ref{journal}). The HEG (resp. MEG) count rates are 0.0061 (resp. 0.014) cts\,s$^{-1}$ at maximum and 0.0046 (resp. 0.0095) cts\,s$^{-1}$ at minimum: the different exposure times thus allow us to have data of similar quality (with $\sim 900$ and 2000 cts for HEG and MEG, respectively) at both phases, facilitating comparisons. 

The data were processed using CIAO v4.8 and CALDB v4.7.0. After the initial pipeline processing (task {\sc chandra\_repro}), the high-resolution spectra of each set were combined using the task {\sc combine\_grating\_spectra}, also adding +1 and $-1$ orders. In addition, for each exposure, the 0th order spectrum was extracted in a circle of radius 10px (corresponding to 5'') around the Simbad position of the target while the associated background was evaluated in the surrounding annulus with an outer radius of 30px. Dedicated response matrices were calculated using the task {\sc specextract}. The spectra and matrices were then combined using the task {\sc combine\_spectra} to get a single spectrum for maximum and one for minimum. At the maximum emission phase, the count rate of the 0th order spectrum amounts to 0.014\,cts\,s$^{-1}$, whereas it is 0.0095\,cts\,s$^{-1}$ at minimum. Further spectral analysis was performed within XSPEC v12.9.0i. Note that, for broad-band fitting, all spectra were grouped to reach a minimum of 10 counts per bin. 

\begin{table}
\caption{Journal of the \ch\ observations, ordered by ObsID, with their associated phase according to the ephemeris of \citet{wad11}.}
\label{journal}
\begin{tabular}{lcccc}
\hline\hline
ObsID & Start\_Date & $\Delta T$ & JD & $\phi$\\ 
& & (ks) & & \\
\hline
\multicolumn{4}{l}{\it MAXIMUM (SeqNum 200975)}\\
16653 &2015-07-04 13:58:11 & 38 &2457208.082 &7.06\\
17489 &2015-05-09 07:52:11 & 44 &2457151.828 &6.96\\
17655 &2015-05-12 20:46:27 & 24 &2457155.366 &6.96\\
17694 &2015-07-12 07:35:41 & 36 &2457215.816 &7.08\\
\multicolumn{4}{l}{\it MINIMUM (SeqNum 200976)}\\
16654 &2016-01-08 14:19:48 & 17 &2457396.097 &7.41\\
16655 &2016-01-07 03:19:19 & 18 &2457394.638 &7.41\\
18743 &2016-02-03 15:29:59 & 55 &2457422.146 &7.46\\
18753 &2016-04-11 16:54:57 & 30 &2457490.205 &7.59\\
18754 &2016-03-26 02:40:46 & 50 &2457473.612 &7.55\\
18821 &2016-04-12 14:28:06 & 27 &2457491.103 &7.59\\
\hline
\end{tabular}
\end{table}

\section{3D MHD Model}

\begin{figure}
\includegraphics[width=7cm]{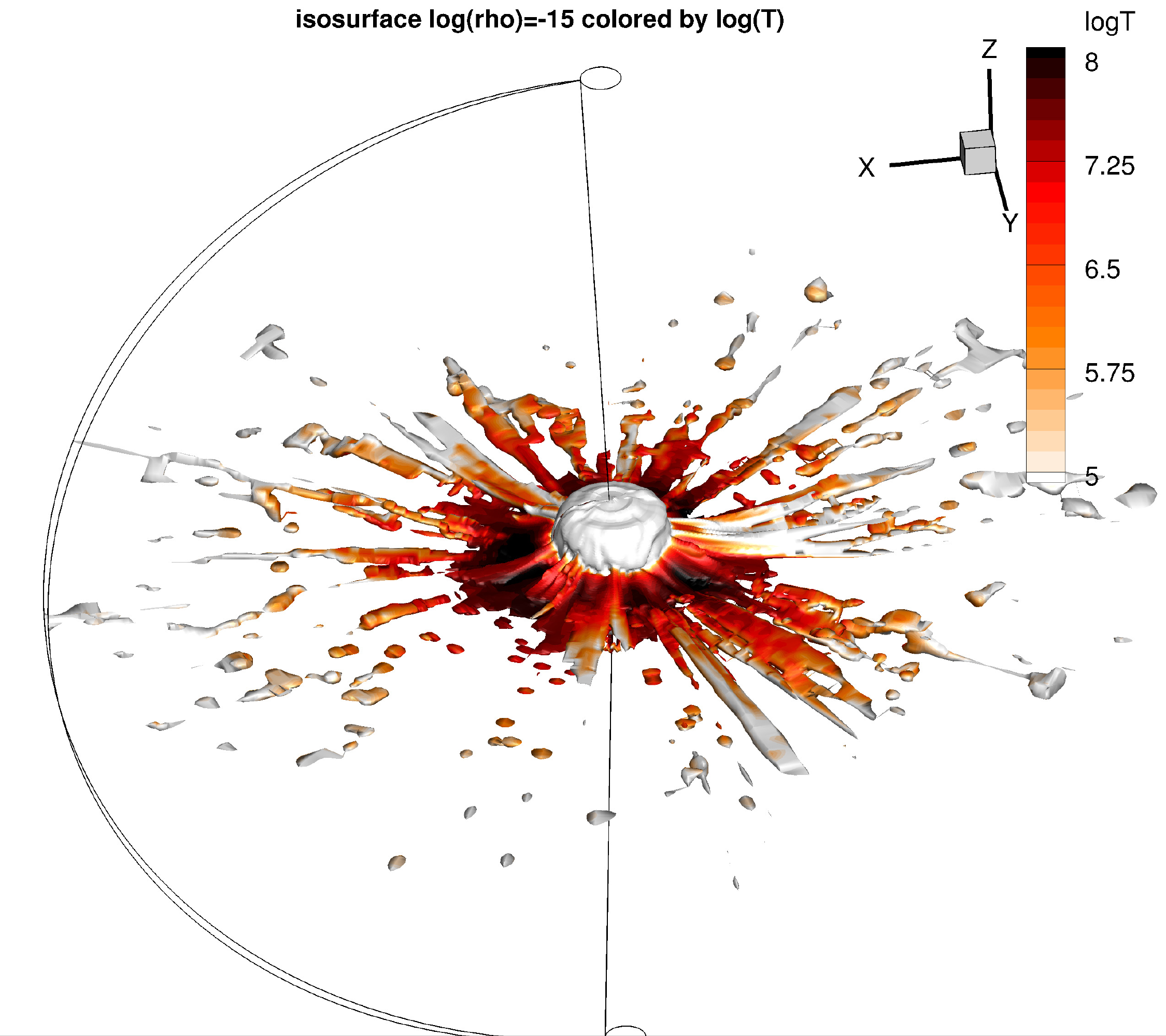}
\caption{A sample view of our 3D MHD model of the star showing an iso-density surface ($\log(\rho)=-15$, with $\rho$ in g\,cm$^{-3}$) colored by logarithm of temperature (in K) at an arbitrary time, t=2\,Ms (end of simulation). Note clearly cool material along the pole, and mostly hot confined wind near the equatorial region. Unlike in 2D MHD models, there is no azimuthal symmetry here, and wind has a range of temperature as visible in this color figure along with numerous scattered dense clouds. Outline of a semi-circle represents the full computational domain extending from 1 to 20 R$_*$, providing a rough scale for comparison.}
\label{model1}
\end{figure}

\begin{figure*}
\includegraphics[width=5.7cm]{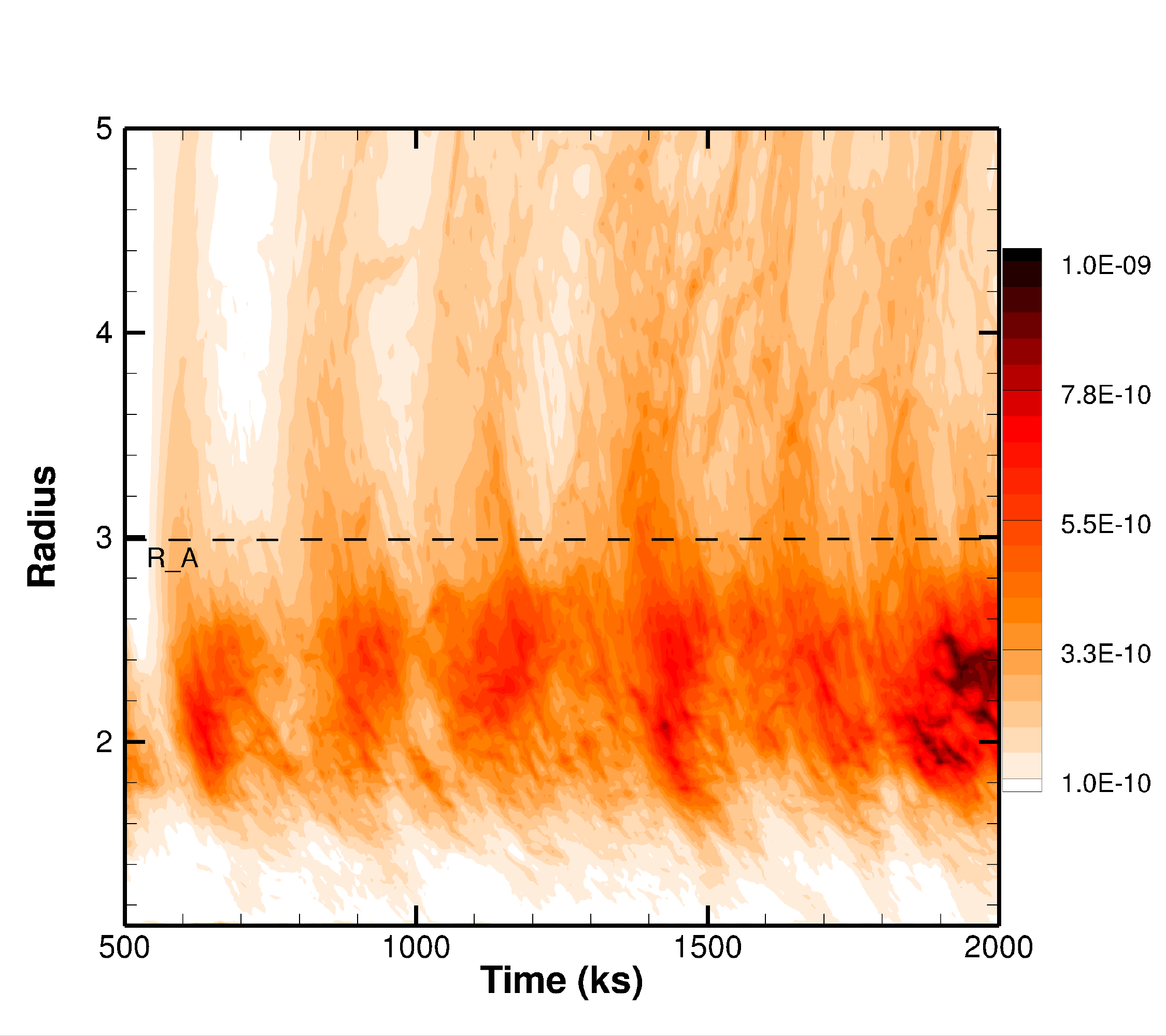}
\includegraphics[width=5.5cm]{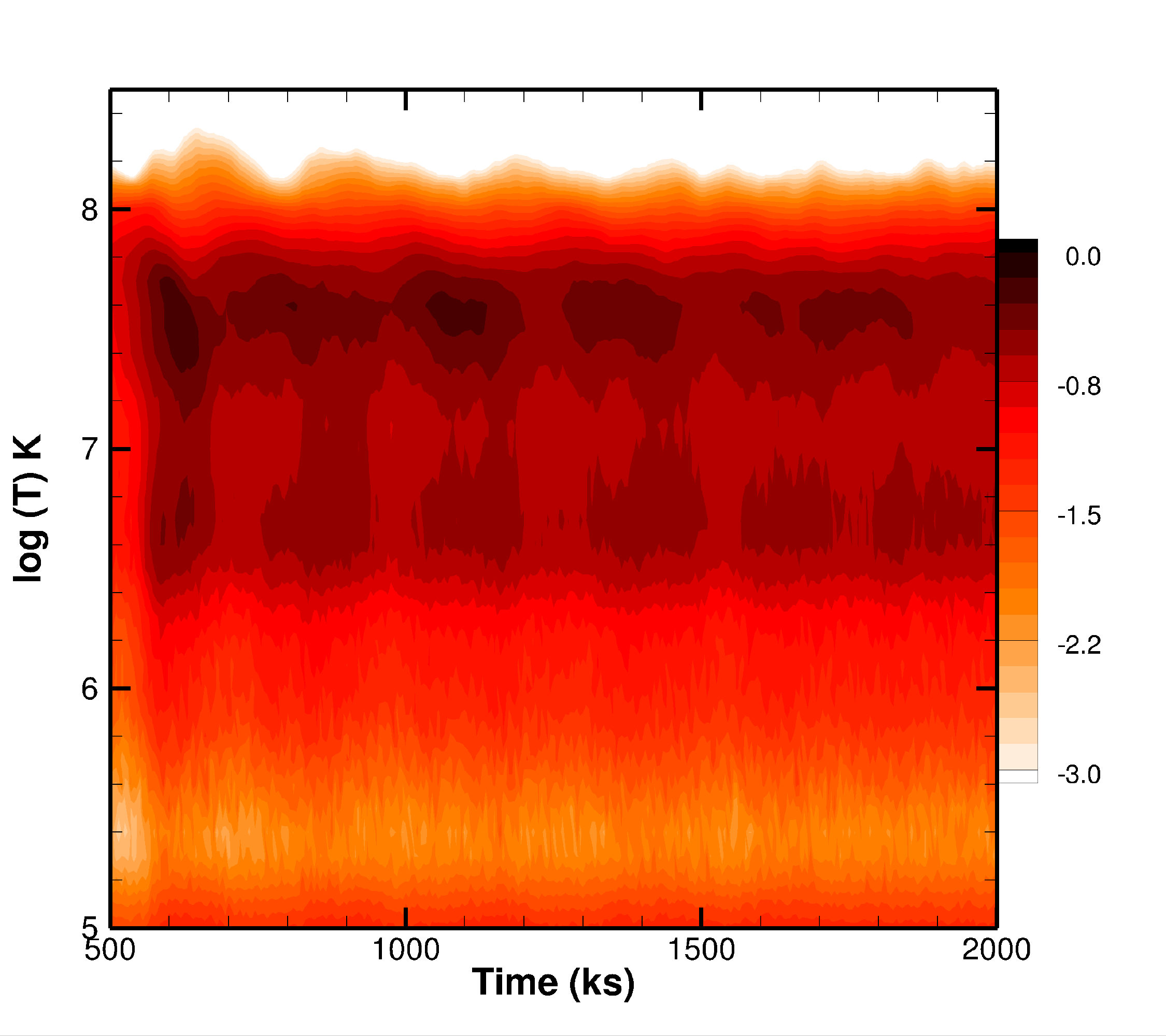}
\includegraphics[width=5.5cm]{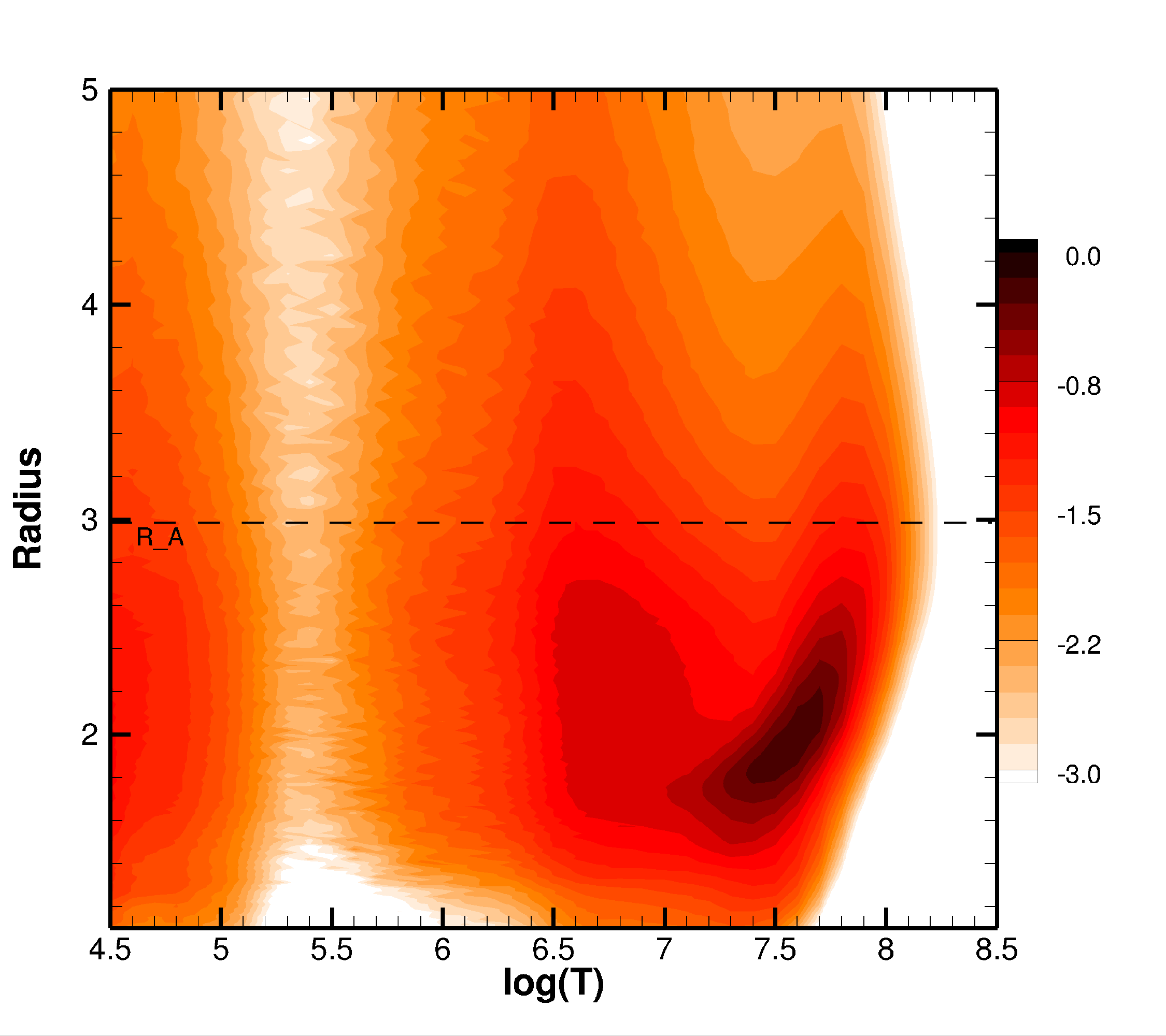}
\caption{{\it Left:} The radial distribution of mass, d$m_e$/d$r$ integrated over full azimuth, plotted versus time and radius (in units of $R_*$). The color bar shows d$m_e$/d$r$ in units of M$_{\odot}$/$R_*$. The horizontal dashed line denotes Alfv\'en radius R$_A$. As evident by a darker band, the magnetosphere is limited within this radius,  with the main plasma component located at 1.6--2.8\,$R_*$.  {\it Middle:}  Distribution of emission measure, $EM(t,T)$, plotted with a logarithmic color scale (normalized to the peak value) versus simulation time (in ks) and logarithm of temperature $\log(T)$ (in K). After an initial transient phase, the plasma distribution appears stable with large volume of gas in the temperature range of 10$^{6.4}$-10$^{7.6}$ K. {\it Right:} Time-averaged distribution of emission measure, $EM(r,T)$, plotted with a logarithmic color scale (normalized to the peak value) versus radius (in units of R$_*$) and logarithm of temperature $\log(T)$ (in K). It demonstrates that  most of the hot and dense gas, source of X-rays, is located within 1.6--2.8\,$R_*$. }
\label{model2}
\end{figure*}

In our procedure to simulate the wind of \hd\ fully self-consistently in 3D, we simply adopted the known stellar parameters of \hd\ \citep[Table \ref{param},][]{sun12}, in order to see how well a detailed, independent simulation of the confined wind based solely on the stellar parameters reproduces X-ray observations. There were no special adjustments to any of the parameters.

\begin{table}
\caption{Stellar parameters used for the MHD model, from \citet{sun12}. }
\label{param}
\begin{tabular}{lc}
\hline\hline 
Parameter & Value\\
\hline
$T_{eff}$ & 35\,kK \\
$\log g$ & 3.5 \\
$R_*$ & 14.5\,R$_{\odot}$\\
$v_{\infty}$ & 2700\,\kms\\
$\dot M$ & $1.6\times10^{-6}$\,M$_{\odot}$\,yr$^{-1}$\\
$B_o$ & 2.45\,kG\\
%$\beta+i$ & 95$^{\circ}$\\
\hline
\end{tabular}
\end{table}

Our basic methods and formalism  for MHD modeling closely follow \citet{udd02} along with \citet{gag05} which includes a detailed energy equation with optically thin radiative cooling \citep{mac81}. The computational grid and boundary conditions are nearly identical to the ones presented in \citet{udd13} for $\theta^1$\,Ori\,C except for the larger extent in radius that now goes from $1R_*$ to $20R_*$ to accommodate for the stronger magnetic field in \hd. %(Fig. \ref{model1}). 
The radiation line force is calculated within the Sobolev approximation using standard CAK \citep{cak75} theory using only the radial component of the force. Since the rotation of \hd\ is extremely slow (period of 537.2\,d, \citealt{wad11}), rotational effects on the dynamics of the wind are expected to be negligible. As such, our model assumes no rotation, and equator throughout this paper will refer to magnetic equator.

Using only adopted stellar parameters, we first relax a non magnetic, spherically symmetric wind model to an asymptotic steady state. This relaxed wind model is then used to initialize density and velocity for our 3D MHD model. For the initial magnetic field, we assume an ideal dipole field with components 
$B_r=B_o (R_* / r)^3 \cos \theta$, 
$B_\theta= (B_o/2) (R_* / r)^3 \sin \theta$, 
and $B_\phi = 0$,
with $B_o$ the polar field strength at the stellar surface.
From this initial condition, the numerical model is then evolved 
forward in time to study the dynamical competition between the field and flow. % (left panel of Fig. \ref{model2}).

The effectiveness of field in channeling wind material depends on its relative strength to wind kinetic energy, and can be characterized by  a dimensionless ``wind magnetic confinement parameter'' \citep{udd02},
\begin{equation}
\eta_\ast \equiv \frac{B_{eq}^2 R_*^2}{{\dot M} v_{\infty}} 
\, ,
\label{etastar}
\end{equation}
where for a dipole, the equatorial field is just half the polar value, $B_{eq} = B_o/2$.
In the case of \hd, $\eta_\ast \approx 50 >> 1$ and the magnetic field dominates the wind outflow near the stellar surface up to a characteristic  {\em Alfv\'{e}n radius}, set approximately by \citet{udd08}:
\begin{equation}
\frac{R_{\rm A}}{R_*} 
\approx 0.3 + \eta_\ast^{1/4}
 \approx 2.95
\, .
\end{equation}
Since the magnetic field energy falls off much more steeply than the wind kinetic energy, the wind can open the field lines above this radius. 

After a short initial transient phase ($<$500\,ks), the simulation settles into a quasi-steady state wherein wind along the poles flow freely whereas material within magnetosphere shocks, cools and then falls back onto the stellar surface in a random fashion, very similar to what happens for the  case of $\theta^1$~Ori~C \citep{udd13}. Fig. \ref{model1} provides a glimpse of this dynamical interaction between the field and the wind: cool polar wind is apparent whereas hot dense material is located around the equator. The clumpiness of the hot confined winds is reminiscent of that observed for their cooler component \citep{sun12,udd13}; it has limited impact on the global properties (X-ray brightness, line profiles,...), as the temporal analysis of the simulations shows. 

To facilitate the understanding of the time evolution of this numerical model, let us follow the approach by \citet{udd13} wherein they define an equatorial radial mass distribution as a function of time:
\begin{equation}
\frac{dm_e}{dr}(r,t)=\frac{1}{2 \pi}\int^{2\pi}_0 \int^{\pi/2+\Delta\theta/2}_{\pi/2-\Delta\theta/2} \rho(r,\theta,\phi,t) \sin \theta \, d\theta \, d\phi 
\end{equation}
where $\Delta \theta=10^\circ$ represents a cone around the equator. The left panel of Fig. \ref{model2} shows this equatorial mass distribution averaged over the azimuth. Clearly, large amount of mass is trapped within the magnetosphere limited by the Alfv\'en radius, $R_A$. Unlike in 2D models where there are clear episodes of emptying and refilling of the magnetosphere, in 3D, on average there is a nearly constant amount of material trapped in the magnetosphere within $1.5-3R_*$.  The middle panel of the same figure shows the differential emission measure (DEM) as a function of simulation time and temperature, demonstrating the near constancy of hot gas, while the right panel shows the DEM but this time as a function of radius and temperature, demonstrating that most of the hot gas is located within the magnetosphere at $\sim 1.6-2.8 R_*$.

\subsection{Predicted X-ray line profiles}

Our fully self-consistent dynamical model allows us to synthesize X-ray line profiles. First, we compute the line-of-sight velocity distribution of the plasma at both phases (pole-on for the maximum emission phase and equator-on for the minimum) as a function of temperature by assuming optically thin wind. To avoid any contamination from initial condition transients, the distributions were time-averaged from 500\,ks to 2000 ks. 

\begin{figure*}
\includegraphics[width=7cm]{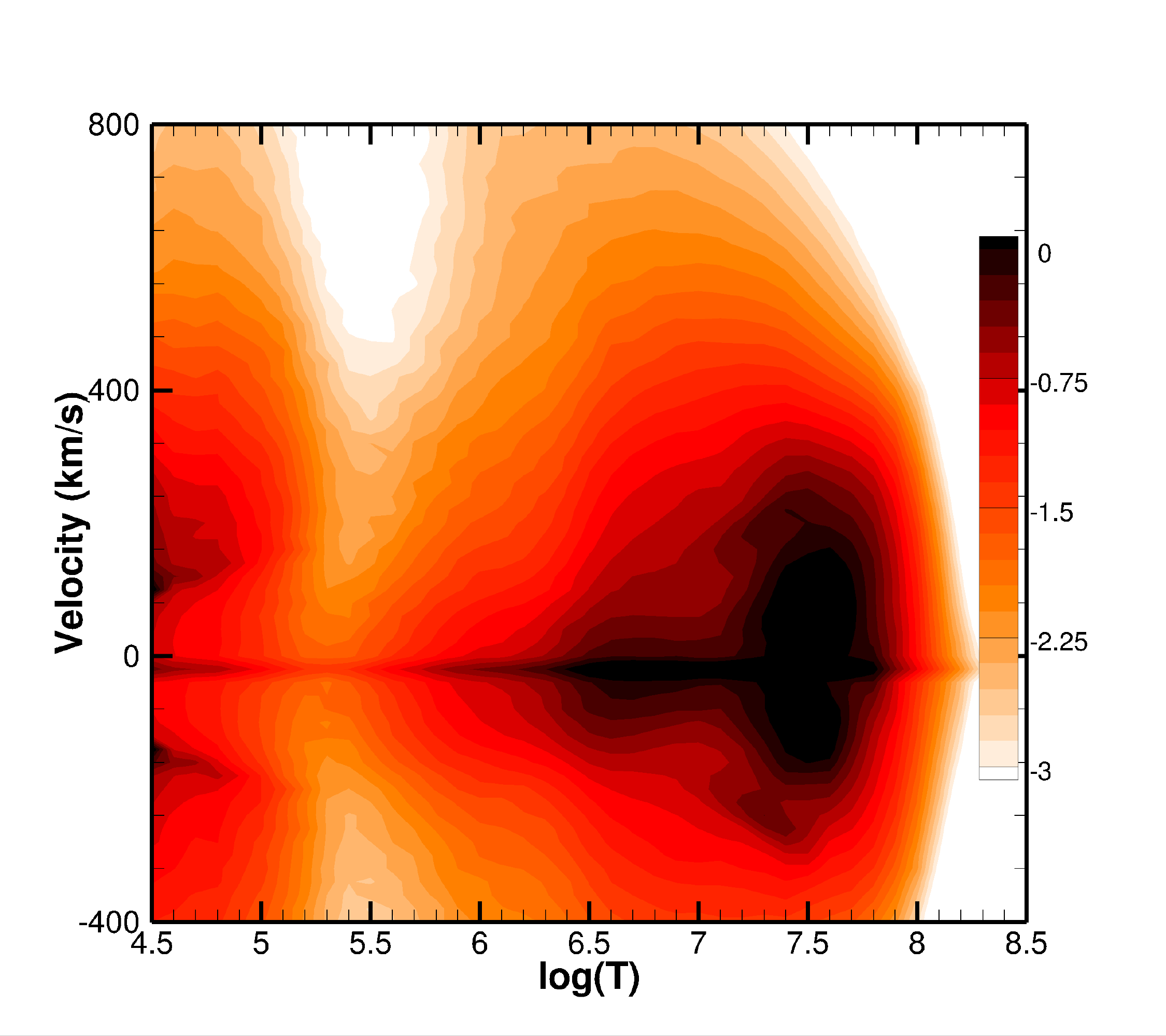}
\includegraphics[width=7cm]{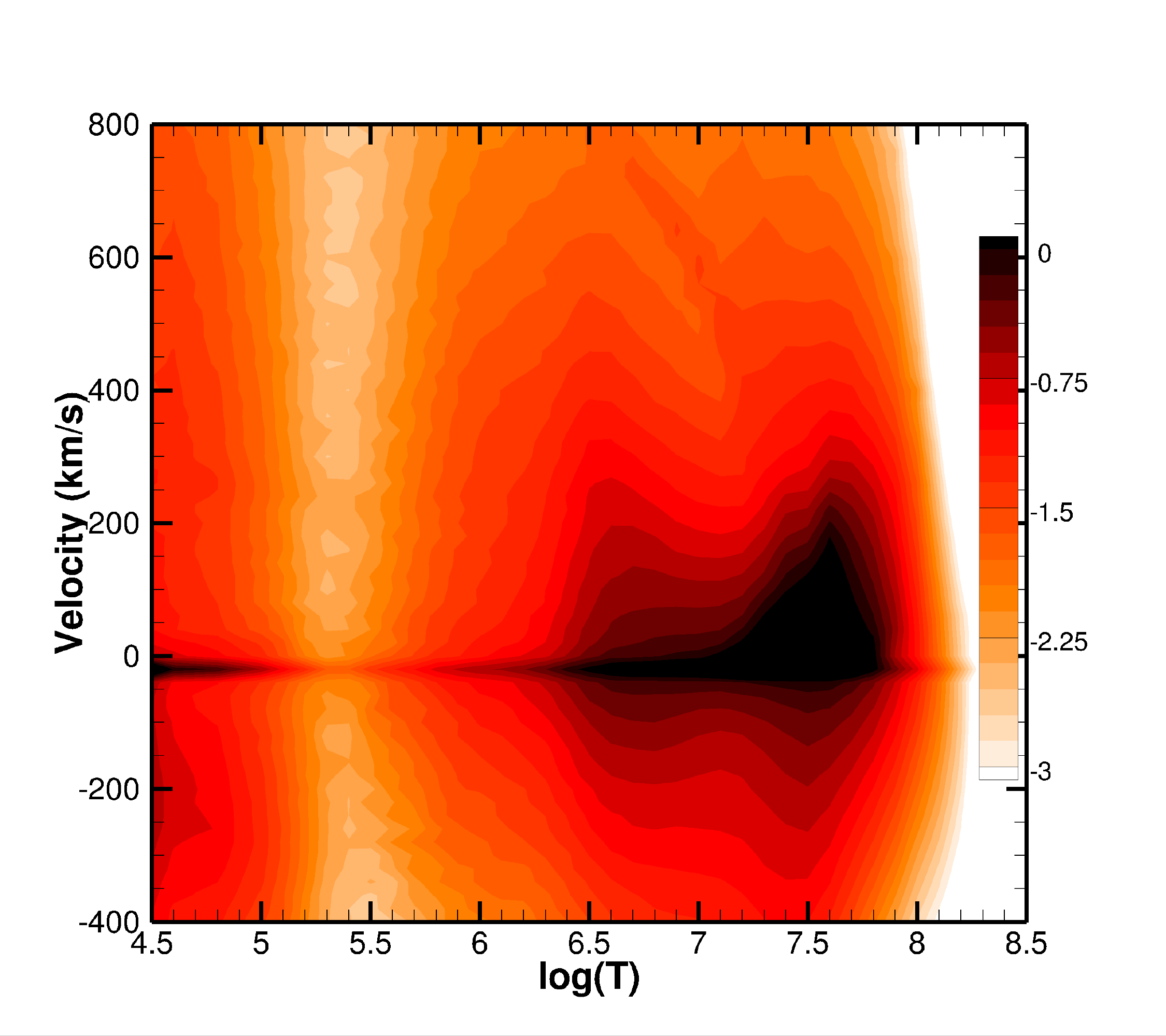}
\caption{The panels show the plasma velocity (in km\,s$^{-1}$) distribution as a function of plasma temperature, with colors representing $EM(v,T)$ in logarithmic units (normalized to the peak value). The left panel represents the maximum emission phase (pole-on view) while the right one corresponds to the equatorial view (minimum emission phase). }
\label{modelline}
\end{figure*}

\begin{figure*}
\includegraphics[width=17cm, bb=25 150 600 520, clip]{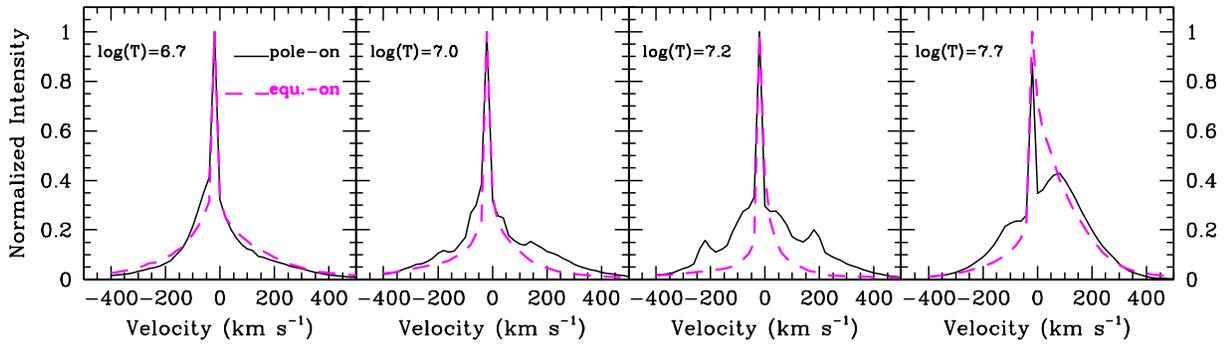}
\caption{Comparison of the simulated line profiles at selected temperatures (pole-on view in black solid line, equator-on view in dashed magenta line). Note that the  emissivity peaks of the Si $fir$ triplet and Lyman$\alpha$ lines occur at $\log(T)=$ 7.0 and 7.2, respectively. }
\label{modelline2}
\end{figure*}

The resulting profiles are shown in Fig. \ref{modelline}, while Fig. \ref{modelline2} compares the line profiles obtained at both phases for selected plasma temperatures. The line shift of the simulated profiles is always close to zero (around $-20$\,\kms) and it is the same at the two phases. Because the plasma is strongly confined near the magnetic equator, the FWHMs always appear quite narrow, about 30\,\kms, but extended wings exist, reaching up to 300\kms\ on each side. These wings reach larger velocities for $\log(T)=7.7$ or at the maximum emission phase. 

\section{Results}

\subsection{Line-by-line analysis of X-ray lines}

\begin{figure}
\includegraphics[width=8.5cm]{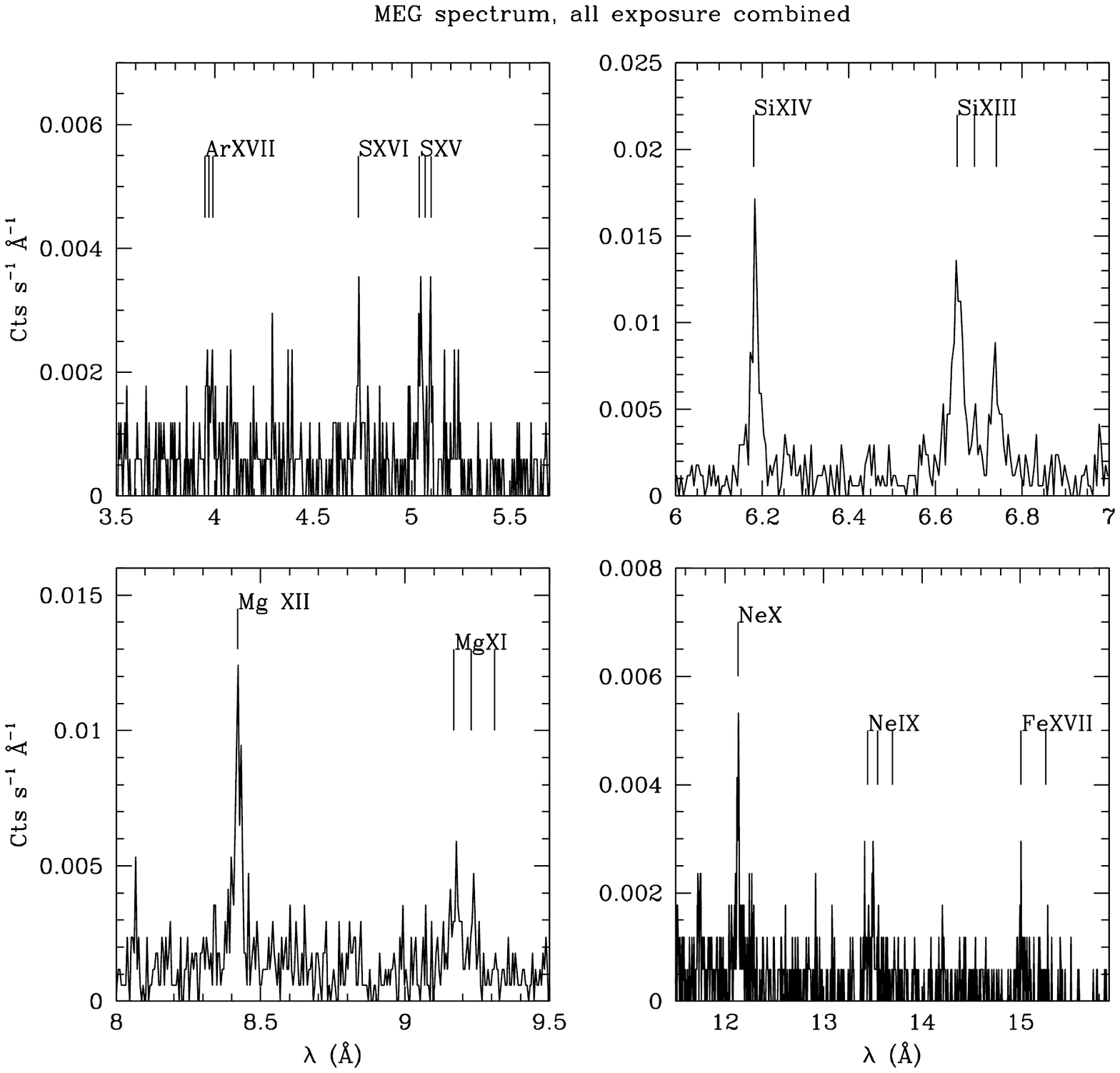}
\caption{MEG spectrum combining all \ch\ exposures, with lines labelled.}
\label{hegmeg}
\end{figure}

%\subsection{Observational characterization}
The high-resolution HEG/MEG spectra reveal the typical lines of massive stars' spectra: $fir$ triplets associated to the He-Like ions of Ar (barely detectable), S, Si, Mg, and Ne, as well as Fe\,{\sc xvii} lines near 15\AA\ and Lyman$\alpha$ lines of H-like S, Si, Mg, and Ne (see Fig. \ref{hegmeg}). The H-like lines of Mg and Si appear stronger than the He-like lines of the same elements, while H-like and He-like lines of S display similar strengths. Such features are untypical for ``normal'' O-stars which rather show very faint or undetectable Mg\,{\sc xii}, Si\,{\sc xiv}, or S\,{\sc xvi} lines, but they were already seen in HD\,148937 and $\theta^1$\,Ori\,C \citep[see in particular their Fig. 2 for a graphical comparison of  X-ray spectra from normal and magnetic O-stars]{naz12}. This underlines the presence of hot plasma in magnetic stars. 

Not all lines have enough counts to provide a meaningful fit, however. Only the strong X-ray lines in the 5--12\AA\ range were fitted by Gaussians, using Cash statistics and unbinned spectra. Fitting was simultaneously performed on both HEG and MEG spectra, to increase signal-to-noise. For Lyman$\alpha$ lines, two Gaussians are used: the two components were forced to share the same velocity and width, and their flux ratio was fixed to the theoretical one in ATOMDB\footnote{See e.g. http://www.atomdb.org/Webguide/webguide.php}. For $fir$ triplets, four Gaussians were used, sharing the same velocity and width, and the flux ratio between the two intercombination lines was again fixed to the theoretical one. No background subtraction was done before fitting: a simple, flat power law being used to represent the local background around the considered lines. Table \ref{linefit} yields the line properties measured for both the maximum and minimum phases, with 1$\sigma$ errors determined using the {\sc error} command under XSPEC. 

\begin{table}
\caption{Properties of the X-ray lines, with 1$\sigma$ errors.}
\label{linefit}
\begin{tabular}{lcc}
\hline\hline
Line & Max & Min \\ 
\hline
\multicolumn{3}{l}{\it Lyman$\alpha$ lines}\\
Si\,{\sc xiv}\\
$v$ (\kms) & 0$\pm$63 & 0$\pm$88 \\
FWHM (\kms) & 399$\pm$300 & 871$\pm$231 \\
$F_x$ (10$^{-6}$ ph\,cm$^{-2}$\,s$^{-1}$) & 4.58$\pm$0.39 & 2.78$\pm$0.34 \\
Mg\,{\sc xii}\\
$v$ (\kms) & 79$\pm$90 & 33$\pm$68 \\
FWHM (\kms) & 351$\pm$224 & 557$\pm$170 \\
$F_x$ (10$^{-6}$ ph\,cm$^{-2}$\,s$^{-1}$) & 3.05$\pm$0.59 & 1.88$\pm$0.34 \\
Ne\,{\sc x}\\
$v$ (\kms) & 68$\pm$123 & $-47\pm$44 \\
FWHM (\kms) & 776$\pm$327 & 52$\pm$233 \\
$F_x$ (10$^{-6}$ ph\,cm$^{-2}$\,s$^{-1}$) & 5.25$\pm$1.28 & 4.25$\pm$0.91 \\
\multicolumn{3}{l}{\it He-like triplets}\\
S\,{\sc xv}\\
$v$ (\kms) & 95$\pm$133 & 181$\pm$158 \\
FWHM (\kms) & 780$\pm$338 & unconstrained \\
$F_x(f)$ (10$^{-6}$ ph\,cm$^{-2}$\,s$^{-1}$) & 2.06$\pm$0.60 & 0.41$\pm$0.30 \\
$F_x(i)$ (10$^{-6}$ ph\,cm$^{-2}$\,s$^{-1}$) & 0.15$\pm$0.42 & 0.51$\pm$0.34 \\
$F_x(r)$ (10$^{-6}$ ph\,cm$^{-2}$\,s$^{-1}$) & 2.66$\pm$0.63 & 0.90$\pm$0.39 \\
$f/i$ & 13.9$\pm$39.3 & 0.80$\pm$0.80 \\
$f+i/r$ & 0.83$\pm$0.34 & 1.02$\pm$0.67 \\
Si\,{\sc xiii}\\
$v$ (\kms) & 100$\pm$67 & $-148\pm$112 \\
FWHM (\kms) & 524$\pm$161 & 1239$\pm$203 \\
$F_x(f)$ (10$^{-6}$ ph\,cm$^{-2}$\,s$^{-1}$) & 1.45$\pm$0.33 & 1.75$\pm$0.34 \\
$F_x(i)$ (10$^{-6}$ ph\,cm$^{-2}$\,s$^{-1}$) & 1.45$\pm$0.37 & 0.79$\pm$0.42 \\
$F_x(r)$ (10$^{-6}$ ph\,cm$^{-2}$\,s$^{-1}$) & 3.91$\pm$0.50 & 3.09$\pm$0.47 \\
$f/i$ & 1.00$\pm$0.34 & 2.22$\pm$1.25 \\
$(f+i)/r$ & 0.74$\pm$0.16 & 0.49$\pm$0.09 \\
\hline
\end{tabular}
\end{table}

No significant line shift is detected, as expected (Figs. \ref{modelline}, \ref{modelline2}) but also as seen in the \xmm\ data of \hd\ \citep{naz07} or in the \ch\ data of $\theta^1$\,Ori\,C \citep{gag05} and HD\,148937 \citep{naz12}. Averaging across the five line sets yields a mean velocity of 68$\pm$44\kms at maximum flux and 4$\pm$46\kms at minimum flux. This slightly larger redshift at maximum, when confined winds are seen face-on, is contrary to what was observed for the average line shift of $\theta^1$\,Ori\,C (where the mean radial velocity changed from $-75$ to 93\kms\ at the same phases) - but the errors are large, requiring confirmation. If this occurs, then it would constitute another difference in behavior between HD\,191612 (and more largely Of?p stars) and $\theta^1$\,Ori\,C, along with the known differences in X-ray hardness and in its variations \citep[the very hard X-rays of $\theta^1$\,Ori\,C somewhat soften while brightening, opposite to the behavior of the softer X-ray emission of HD\,191612][]{naz07,naz14}, as well as in UV lines \citep[opposite behavior in C\,{\sc iv} and N\,{\sc v}, see e.g.][]{mar13,naz15}: it might help us understand the physical origin of this (still unexplained) difference.

The X-ray lines detected in HETG generally appear resolved, with FWHMs of 400--1200\kms. The observed profiles indicate broader FWHMs than simulations (Figs. \ref{modelline}, \ref{modelline2}), a problem already encountered for $\theta^1$\,Ori\,C \citep{gag05} though with a lesser amplitude. Observed lines also appear slightly broader at the minimum phase (changing from 566$\pm$124\kms\ to 680$\pm$105\kms\ on average between the two phases) while the predicted wings of the simulated profiles instead appeared broader at maximum phase, but the difference only amounts to 1--2$\sigma$ and is thus only marginal. We however come back to this issue in the direct comparison section. It must finally be noted that, in the \xmm-RGS data of \hd, significantly larger FWHMs $\sim$2000\kms were found for the lower-Z lines. As in HD\,148937 and $\theta^1$\,Ori\,C, we thus find both narrow and broad lines in the X-ray spectrum, pointing to a mixed origin of the X-ray emitting plasma. 

Of course, line fluxes change between the maximum and minimum emission phase: when the overall flux change is about 40\% (see next section), the line fluxes accordingly vary by 20--60\%. Furthermore, the $(f+i)/r$ ratios slightly increase at minimum emission phase while the H-to-He like flux ratio of Si slightly decrease in parallel: this marks a slight decrease in plasma temperature at minimum flux, in agreement with the overall softening of the broad-band spectra already reported by \citet[see also next subsection]{naz07,naz14}. On the other hand, no clear, significant variation of the $f/i$ ratios can be detected, but they are affected by large errors prohibiting detection of all but extremely large changes. 

To reach more quantitative results, we focus on the Si lines as they have the lowest uncertainties. Assuming the $f/i$ ratio does not change much, we then combine all high-resolution spectra, i.e. minimum and maximum phases together, and perform a similar line analysis as just described, deriving a value of 1.46$\pm$0.39 for this $f/i$ ratio. Correcting for the interstellar absorption of $3.2\times10^{21}$\,cm$^{-2}$ is unnecessary for ratios involving the closely-spaced $fir$ lines, but such a correction (by a factor 0.96) needs to be performed for the H-to-He like flux ratios since the lines are more distant in this case. Following the method in \citet{naz12} considering the stellar parameters from Table \ref{param} ($T_{eff}$=35kK and $\log(g)$=3.5, see also \citealt{wad11}), we then draw the following conclusions. 

First, the plasma temperature $\log(T)$ amounts to 7.1$\pm$0.3 at maximum and 6.96$\pm$0.26 at minimum following the triplet ratios, or 7.12$\pm$0.02 and 7.08$\pm$0.03, respectively, considering the H-to-He like flux ratios.  This corresponds to a temperature of $\sim$1\,keV with a small decrease (by 25\%, which is about $1\sigma$) between maximum and minimum. This agrees well with results derived from global fits (see temperature and hardness ratios in next subsection) but it is lower than the typical plasma temperature in the model (see middle panel of Fig. \ref{model2}). 

Second, the initial formation radius, derived from the spectra combining all exposures, is at 2.4$\pm$0.7\,$R_*$ (it is at 1.7$\pm$0.4\,$R_*$ considering the maximum spectra only). This value is similar to the formation radius derived for $\theta^1$\,Ori\,C \citep[1.6--2.1\,$R_*$, see][erratum]{gag05} and HD\,148937 \citep[1.9$\pm$0.4\,$R_*$, see][]{naz12}. It also correlates well with the position of the hot plasma in our 3D MHD simulation (Fig. \ref{model2}).

\subsection{Observational characterization of the low-resolution spectra}

\begin{figure}
\includegraphics[width=8cm]{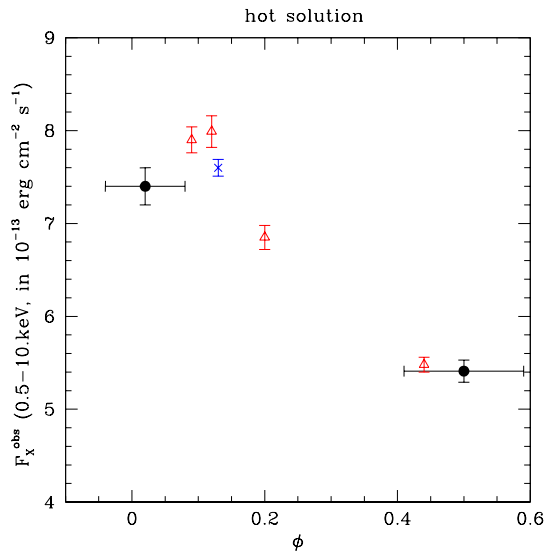}
\caption{Evolution of the observed X-ray flux with phase, from the ``hot''solution of the fits (see Table \ref{globalfit}). The \ch\ values are presented using black dots, the horizontal error bar corresponding to the phase span of the observations (Table \ref{journal}); the four \xmm\ datasets from 2005 are shown with open red triangles and that from 2008 with a blue cross. The agreement between both observatories is remarkable, considering the remaining differences in instrumental calibrations. It demonstrates the repetitive behavior of \hd\ at high energies.}
\label{globalfittime}
\end{figure}

The 0th order spectra of \hd\ can be well fitted with two absorbed thermal components. Beyond the interstellar absorption ($3.2\times10^{21}$\,cm$^{-2}$, \citealt{naz14}), an additional absorption can be allowed, considering the presence of circumstellar material. This absorption can be either added to each thermal component (as in \citealt{naz07}) or be of a global nature (as in \citealt{naz14}). In this paper, we choose the latter option as there is no need for an additional degree-of-freedom when fitting the \ch\ 0th spectra. Besides, the usual trade-off between a hot plasma with little additional absorption and a warm plasma with more absorption is again found: as they represent equally well the data and no prior knowledge favors one possibility over the other, both solutions are provided in Table \ref{globalfit}. Note that, as the additional absorption or the temperature did not significantly change across the fits, we fixed them and it is the results of these constrained fits which are shown in the table. These fits were performed assuming the solar abundance of \citet{asp}, which is why we also provide new fits for the \xmm\ spectra previously presented in \citet{naz07,naz14}. \hd, as other Of?p stars, appears slightly enriched in nitrogen and depleted in carbon and oxygen \citep{mar15}. However, considering non-solar abundances (either by fixing them to the Martins et al. values or letting them vary freely) does not significantly improve the quality of the fits, nor does it change the conclusions, so we kept them solar. 

The new \ch\ data confirm the results derived previously on \xmm\ observations: the flux of \hd\ increase by $\sim$40\% at maximum emission phase, and the X-ray emission appears harder when brighter. This strong agreement (see also Fig. \ref{globalfittime}) demonstrates the great stability in the X-ray properties of \hd\ over a decade (i.e. 7 periods of \hd). 

\begin{sidewaystable}
\footnotesize
\caption{Best-fit spectral parameters for a model of the type $wabs \times phabs \times \sum_1^2 apec$. The columns $F_{\rm X}^{unabs}$ provide fluxes corrected for the interstellar absorbing column ($3.2\times10^{21}$\,cm$^{-2}$) only, while $F_{\rm X}^{int}$ yield the fluxes corrected for the full absorbing columns The hardness ratios $HR$ are defined as $F_{\rm X}^{unabs}$(hard)/$F_{\rm X}^{unabs}$(soft), with ``tot'', ``soft'', and ``hard'' corresponding to the 0.5--10.\,keV, 0.5--2.\,keV, and 2.--10.\,keV energy bands, respectively. The 1$\sigma$ errors were calculated in XSPEC using the {\sc error} command for parameters and the {\sc flux err} command for observed fluxes (the errors on $HR$s were calculated assuming the relative errors on observed and absorption-corrected fluxes are equal). If asymmetric errors were found, the largest value is shown here.}
\label{globalfit}
\begin{tabular}{lcccccccccccc}
\hline\hline
ID & $\phi$ & $norm_1$ & $norm_2$ & $\chi^2_{red}$ (dof) & $F_{\rm X}^{obs}$ (tot) & $F_{\rm X}^{obs}$ (soft) & $F_{\rm X}^{obs}$ (hard) & \multicolumn{3}{c}{$F_{\rm X}^{unabs}$ (tot, soft,hard)}& $F_{\rm X}^{int}$ (tot) & $HR$\\
& & ($10^{-4}$\,cm$^{-5}$) & ($10^{-4}$\,cm$^{-5}$) & & \multicolumn{7}{c}{($10^{-13}$\,erg\,cm$^{-2}$\,s$^{-1}$)} & \\ 
\hline
\multicolumn{12}{l}{\it ``Warm'' solution: $N_{\rm H}^{add}=5\times10^{21}$\,cm$^{-2}$, $kT$=0.24 and 1.8\,keV} \\
\ch\ 0th (200975)& 0.02 & 54.0$\pm$3.2 &7.66$\pm$0.26 &1.14(140) &7.38$\pm$0.19 &4.56$\pm$0.16 &2.82$\pm$0.09 &14.2 &11.2 &2.97 &61.3 &0.265$\pm$0.013\\
\ch\ 0th (200976)& 0.50 & 47.2$\pm$2.6 &5.21$\pm$0.18 &0.98(149) &5.53$\pm$0.17 &3.61$\pm$0.12 &1.92$\pm$0.06 &11.2 & 9.2 &2.03 &51.6 &0.221$\pm$0.010\\
XMM (0300600201) & 0.09 & 55.9$\pm$1.5 &8.01$\pm$0.23 &1.51(169) &7.66$\pm$0.13 &4.74$\pm$0.09 &2.92$\pm$0.11 &14.6 &11.6 &3.07 &62.9 &0.265$\pm$0.011\\
XMM (0300600301) & 0.20 & 53.1$\pm$1.2 &6.63$\pm$0.20 &1.66(168) &6.68$\pm$0.12 &4.25$\pm$0.06 &2.42$\pm$0.09 &13.1 &10.6 &2.55 &58.5 &0.241$\pm$0.010\\
XMM (0300600401) & 0.44 & 45.9$\pm$7.8 &5.03$\pm$0.11 &1.76(236) &5.34$\pm$0.06 &3.50$\pm$0.05 &1.84$\pm$0.06 &10.8 & 8.9 &1.93 &49.7 &0.217$\pm$0.008\\
XMM (0300600501) & 0.12 & 60.0$\pm$1.5 &7.85$\pm$0.26 &1.91(150) &7.76$\pm$0.14 &4.89$\pm$0.06 &2.86$\pm$0.13 &15.1 &12.1 &3.01 &66.6 &0.249$\pm$0.012\\
XMM (0500680201) & 0.13 & 56.4$\pm$1.0 &7.50$\pm$0.14 &1.90(241) &7.38$\pm$0.09 &4.64$\pm$0.08 &2.73$\pm$0.08 &14.3 &11.5 &2.88 &62.8 &0.250$\pm$0.008\\
\multicolumn{12}{l}{\it ``Hot'' solution: $N_{\rm H}^{add}=0$, $kT$=0.75 and 2.4\,keV} \\           
\ch\ 0th (200975)& 0.02 &3.14$\pm$0.19 &5.06$\pm$0.21 &1.12(140) &7.40$\pm$0.20 &4.46$\pm$0.12 &2.94$\pm$0.10 &13.3 &10.2 &3.09 &13.3 &0.303$\pm$0.013\\
\ch\ 0th (200975)& 0.50 &2.56$\pm$0.15 &3.43$\pm$0.15 &1.05(149) &5.41$\pm$0.12 &3.38$\pm$0.11 &2.03$\pm$0.09 &10.0 & 7.8 &2.13 &10.0 &0.273$\pm$0.015\\
XMM (0300600201) & 0.09 &3.24$\pm$0.10 &5.52$\pm$0.20 &1.41(169) &7.90$\pm$0.14 &4.72$\pm$0.10 &3.17$\pm$0.11 &14.1 &10.8 &3.33 &14.1 &0.308$\pm$0.013\\
XMM (0300600301) & 0.20 &3.10$\pm$0.09 &4.50$\pm$0.17 &1.51(168) &6.85$\pm$0.13 &4.23$\pm$0.07 &2.62$\pm$0.10 &12.5 & 9.7 &2.75 &12.5 &0.284$\pm$0.012\\
XMM (0300600401) & 0.44 &2.71$\pm$0.05 &3.37$\pm$0.09 &1.66(236) &5.48$\pm$0.08 &3.49$\pm$0.05 &1.99$\pm$0.06 &10.2 & 8.1 &2.09 &10.2 &0.258$\pm$0.009\\
XMM (0300600501) & 0.12 &3.53$\pm$0.10 &5.34$\pm$0.22 &1.28(150) &7.99$\pm$0.17 &4.90$\pm$0.09 &3.10$\pm$0.13 &14.5 &11.2 &3.25 &14.5 &0.290$\pm$0.013\\
XMM (0500680201) & 0.13 &3.39$\pm$0.68 &5.04$\pm$0.12 &1.38(241) &7.60$\pm$0.09 &4.67$\pm$0.06 &2.93$\pm$0.08 &13.8 &10.7 &3.07 &13.8 &0.287$\pm$0.009\\
\hline
\end{tabular}
\end{sidewaystable}

\subsection{Direct comparison with model predictions}

\begin{figure}
\includegraphics[width=8cm]{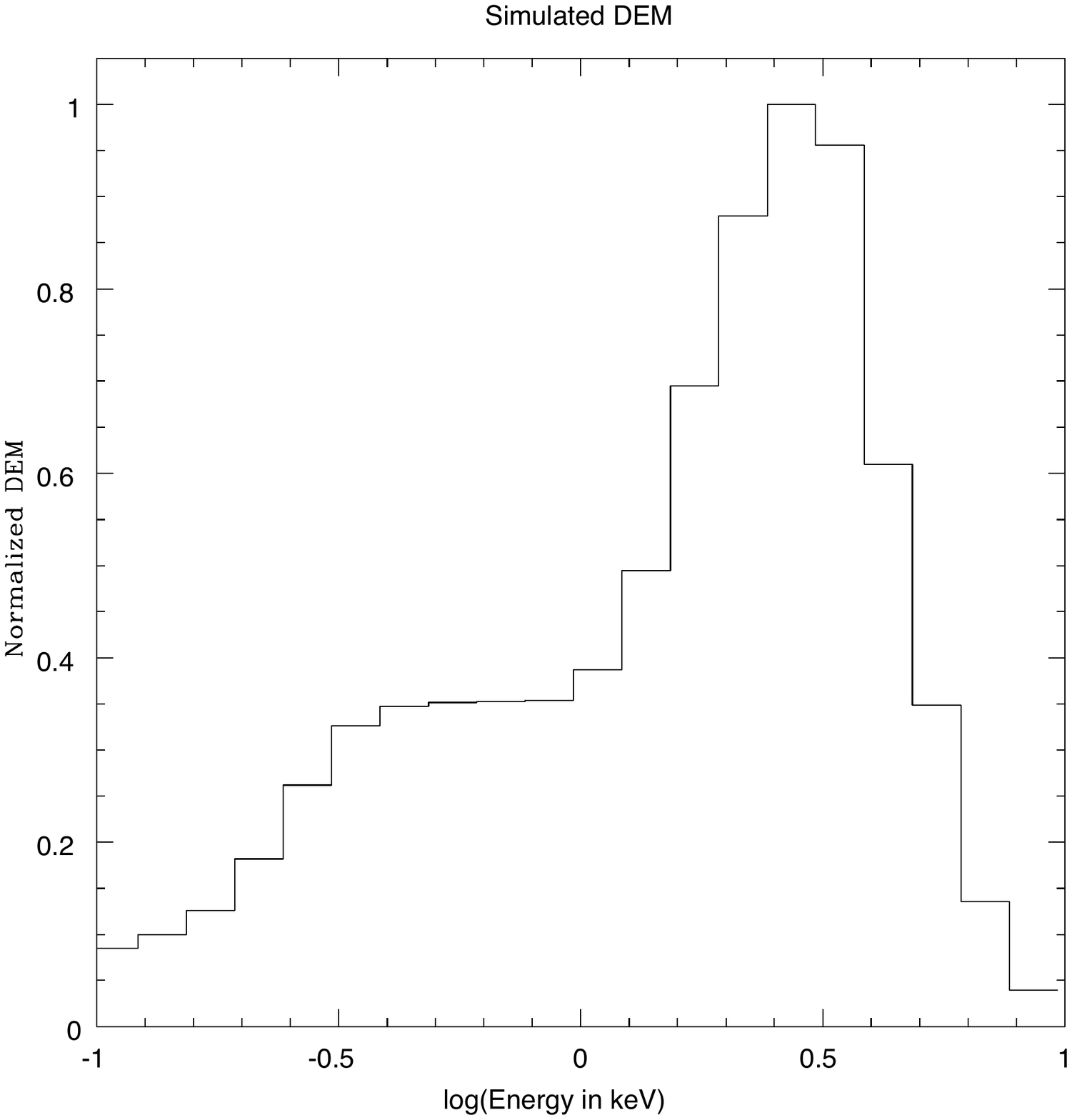}
\caption{The time-averaged DEM per $\log(E)$= 0.1 (in keV) as derived from 3D MHD simulation, normalized to its peak. The total $EM =2.7 \times 10^{56}$\,cm$^{-3}$ when integrated over $\log(T)=$6 to 8, is comparable to what was found in simulations of $\theta^1$\,Ori\,C ($\sim 9 \times 10^{55}$\,cm $^{-3}$ integrated over the same temperature range) though it is somewhat larger as it corresponds to a larger magnetosphere.}
\label{dem}
\end{figure}

In order to make a {\it direct} comparison between the MHD model and observations, we developed a new spectral model for XSPEC. We note that the X-ray emission from the confined winds is thermal and due to the high plasma densities the non-equilibrium ionization effects can be neglected. We thus consider thermal plasma in collisional ionization equilibrium. The model reads in the DEM as provided by the 3D MHD simulations averaged { over 1.5\,Ms of simulation time (from 0.5 to 2.0\,Ms) and over all azimuthal angles} (Fig. \ref{dem}). To calculate the theoretical spectrum associated with it, we make use of the optically thin plasma model ($apec$) for each plasma temperature of the input DEM. As free parameter, the model scaling factor $sc$ indicates whether the total amount of hot plasma as derived in the hydrodynamic simulations (emission measure $=2.7\times 10^{56}$\,cm$^{-3}$ over $\log(T)=$ 6 to 8) matches that required by observation. For example, a $sc = 1$ indicates a perfect correspondence, while $sc < 1$~or $> 1$ means that the MHD model correspondingly predicts higher or smaller amount of hot plasma than required by the data, respectively. Note that abundances of the hot plasma are additional possible free parameters of this model. Finally, our new XSPEC model is able to take into account the kinematic information provided from the 3D MHD simulations as well. Thus, it is able to model the realistic line profiles (more on that further below).

\begin{figure*}
\includegraphics[width=8cm]{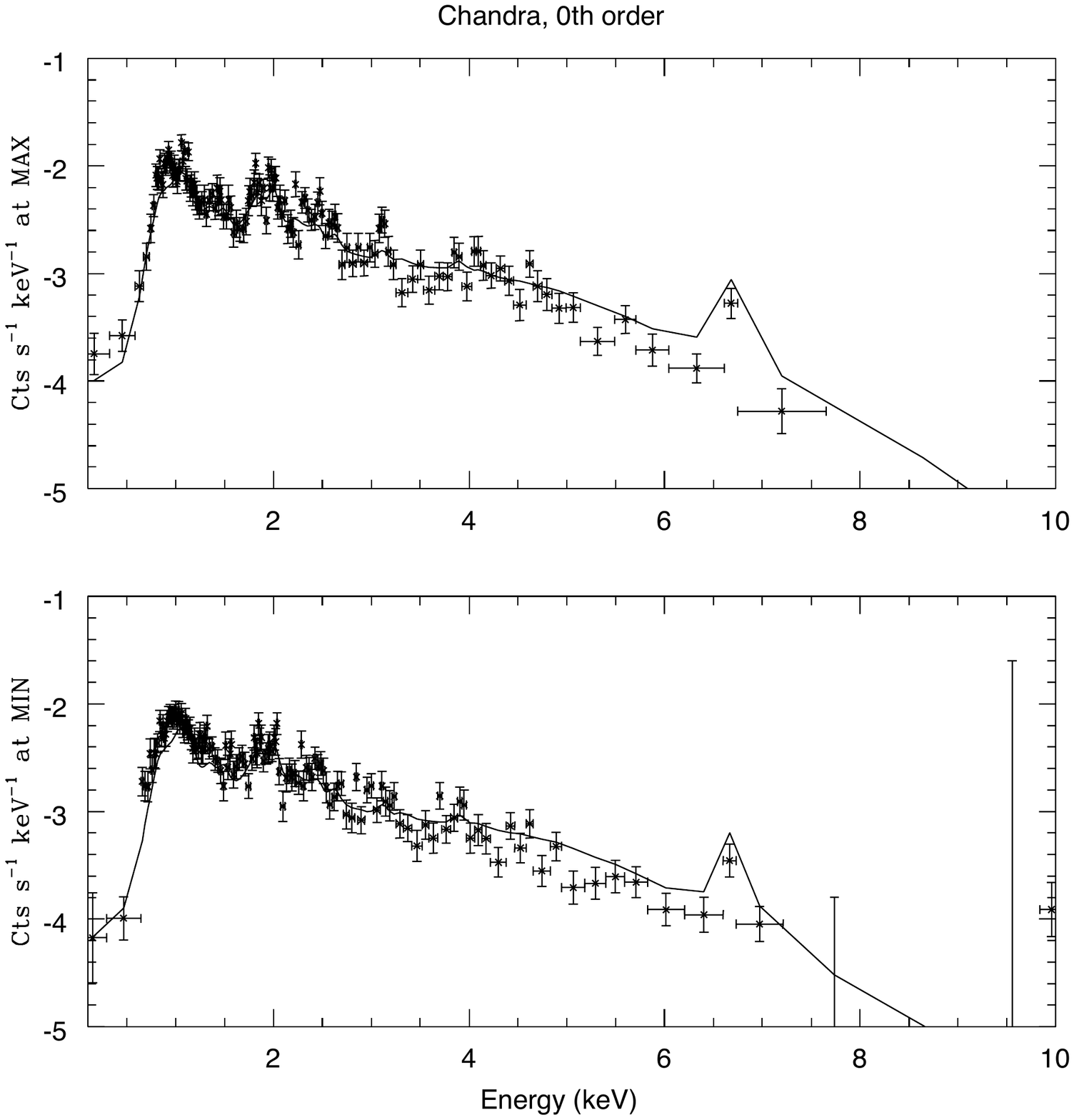}
\includegraphics[width=8cm]{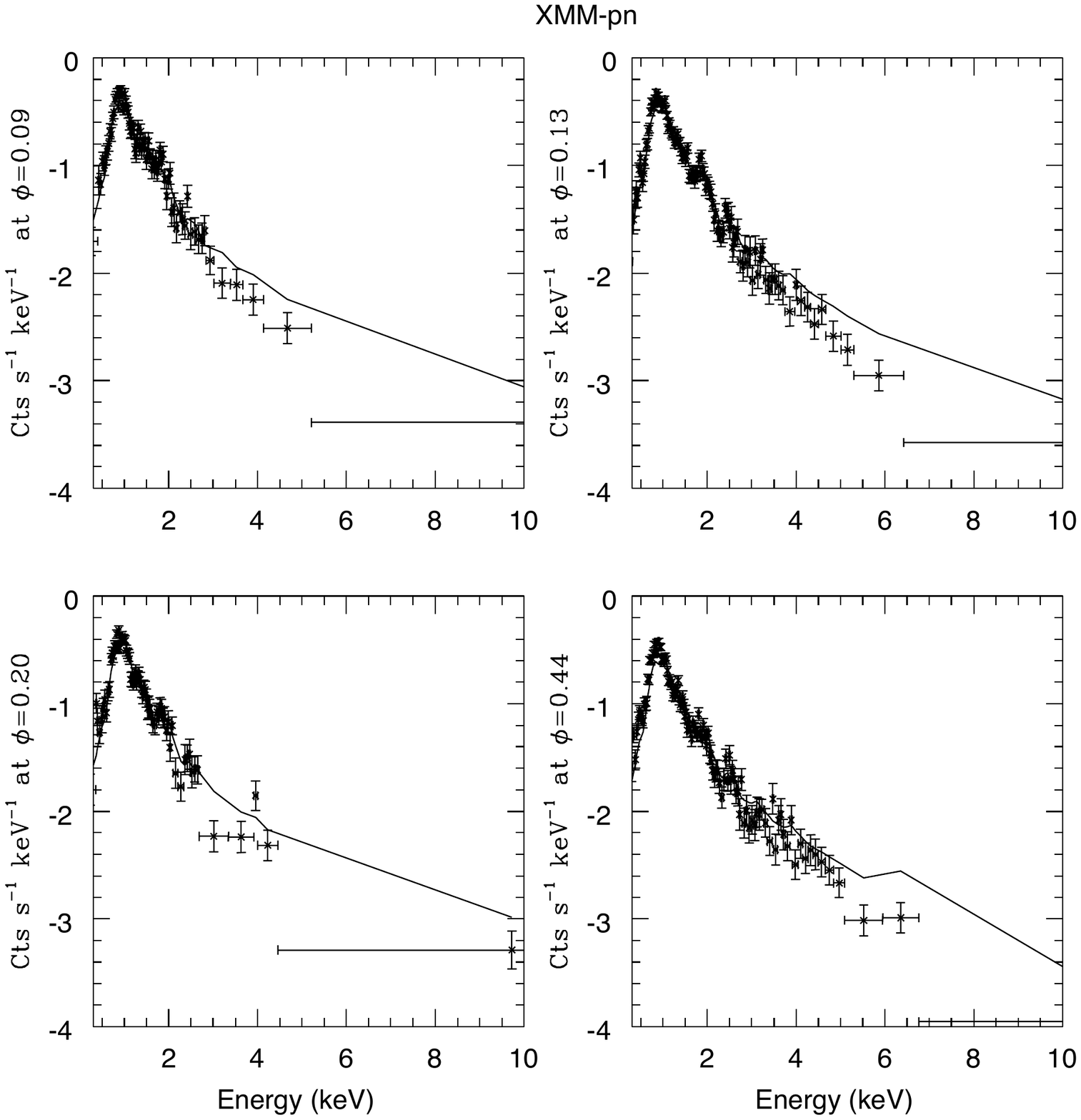}
\caption{Comparison of the best-fit model using the simulated DEM (black line) with \ch\ 0th order spectra (left, for maximum and minimum emission phases) and with \xmm-EPIC pn spectra (right, for four different phases). Note the good fit up to 2--4\,keV, and the slight excess of hard flux at larger energies.}
\label{demcompa1}
\end{figure*}

\begin{figure}
\includegraphics[width=8cm]{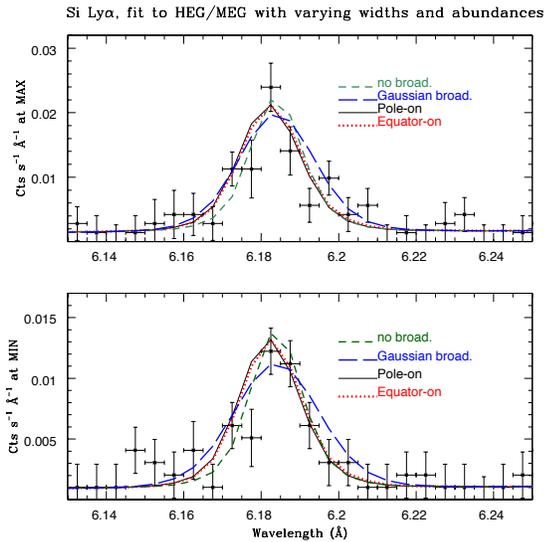}
\caption{The Si\,{\sc xiv} Lyman$\alpha$ line  profile observed in unbinned MEG spectra at maximum (top) and minimum (bottom) emission phases, compared to results of different fits (see Table \ref{demtablevel}): simulated DEM without broadening (green dashed line), with Gaussian broadening (blue long-dashed line), or with the line profile found in simulations (either for a pole-on situation, black solid line, or an equatorial view, dotted red line). }
\label{demcompa2}
\end{figure}

We began by fitting this model to the low-resolution spectra (both 0th order \ch\ and \xmm\ spectra). First, we allowed the possibility of absorption in addition to the interstellar column ($3.2\times10^{21}$\,cm$^{-2}$, see above), but this results in a 1$\sigma$ upper limit on $N_{\rm H}^{add}$ of $2\times10^{19}$\,cm$^{-2}$, indicating that the interstellar absorption is sufficient to fit the spectra. We therefore consider only the (fixed) interstellar absorbing column in what follows. The results of the fits are provided in Table \ref{demtable} and shown in Fig. \ref{demcompa1}. Again, a very good agreement is found between \xmm\ and \ch\ results. The spectra appear very well fitted up to 3\,keV, but the model slightly overpredicts the flux at higher energies. This is reflected in the hardenss ratio $HR$, which is about 0.45 (fixed value, since the DEM shape is fixed) when simpler fits favor values of $\sim$0.25--0.3 (Table \ref{globalfit}). This can be explained by the presence of plasma at high temperatures in the MHD model (see previous sections, in particular Fig. \ref{model2}). 

The scaling factors also indicate an overprediction of the X-ray output by a factor $\sim5$. The added third dimension is a bit less efficient than 2D, but one other suggestion to explain this difference is the chosen value of the mass-loss rate. Indeed, considering the dense wind of \hd, cooling should be efficient, hence $L_{\rm X}\sim \dot M^{-2}$, and it is known that mass-loss rates of massive stars are overestimated by a factor $\sim3$ because of clumping within the stellar wind \citep[e.g.][]{bou05}. In addition, such a reduction of mass loss rate would lead to an effect called `shock retreat' \citep{udd14}, wherein shocked gas retreats towards the stellar surface along the field lines where the velocities are lower, leading to lower shock speeds hence possibly to softer X-rays - though the effect needs to be quantified exactly for a star like \hd. However, we should keep in mind that even the dynamical model presented here has its own shortcomings, e.g. it only uses the radial component of the radiative force and it ignores cooling due to inverse Compton scattering which, although having a relatively minor effect for O stars \citep{udd14}, does slightly reduce the amount of {\it hard} X-rays. Future models should indeed address these shortcomings.

As the simulated DEM is an average value, it is not made to reproduce the flux variations recorded for \hd. Such variations are usually considered to be due to occultation of the hot plasma by the stellar body, when confined winds are seen edge-on. However, such a simple occultation cannot match the observed decrease in flux considering the hot plasma location (about 1.6--2.8\,$R_*$ for \hd, see previous sections): at this position, occultation effects would lead to flux changes of about 15\%. To get the observed 40\% would require an improbably close location for the confined winds \citep[$r<1.2R_*$, see][]{udd16}. Therefore, an additional mechanism is needed. A plausible scenario is the presence of asymmetries in the confined wind structure, which would enhance occultation effects. They could be linked e.g. to an off-center magnetic dipole or to multipolar components to the magnetic field. Current spectropolarimetric observations only sample the dipolar component of the magnetic field, yielding no constraint yet on such features. More precise knowledge of the magnetic geometry, and its consequences on the wind confinement through a new modeling, is thus needed before the increased flux variability can be understood.

\begin{table*}
\footnotesize
\caption{Best-fit parameters for the dedicated confined wind model for low-resolution spectra. }
\label{demtable}
\begin{tabular}{lccccccc}
\hline\hline
ID & $\phi$ & $sc$ & $\chi^2_{red}$ (dof) & $F_{\rm X}^{obs}$ (tot) & $F_{\rm X}^{obs}$ (soft) & $F_{\rm X}^{obs}$ (hard) & $F_{\rm X}^{unabs}$ (tot) \\
& &  & & \multicolumn{4}{c}{($10^{-13}$\,erg\,cm$^{-2}$\,s$^{-1}$)} \\ 
\hline
\ch\ 0th (200975)& 0.02 & 0.202$\pm$0.005 &1.68(141) &7.34$\pm$0.20 &3.72$\pm$0.10 &3.61$\pm$0.08 & 12.5\\
\ch\ 0th (200976)& 0.50 & 0.144$\pm$0.004 &1.69(150) &5.24$\pm$0.13 &2.66$\pm$0.07 &2.58$\pm$0.07 & 8.91\\
XMM (0300600201) & 0.09 & 0.247$\pm$0.004 &1.77(170) &8.97$\pm$0.13 &4.57$\pm$0.09 &4.40$\pm$0.09 & 15.2\\
XMM (0300600301) & 0.20 & 0.222$\pm$0.003 &2.08(169) &8.06$\pm$0.12 &4.11$\pm$0.07 &3.95$\pm$0.07 & 13.7\\
XMM (0300600401) & 0.44 & 0.174$\pm$0.002 &3.11(237) &6.31$\pm$0.08 &3.22$\pm$0.05 &3.09$\pm$0.05 & 10.7\\
XMM (0300600501) & 0.12 & 0.263$\pm$0.004 &1.85(151) &9.55$\pm$0.14 &4.86$\pm$0.10 &4.68$\pm$0.08 & 16.3\\
XMM (0500680201) & 0.13 & 0.237$\pm$0.002 &2.58(242) &8.59$\pm$0.09 &4.38$\pm$0.06 &4.21$\pm$0.06 & 14.6\\
\hline
\end{tabular}
\end{table*}

\begin{table*}
\footnotesize
\caption{Best-fit parameters for the dedicated confined wind model for high-resolution \ch\ spectra. Abundances are in number, relative to Hydrogen and with respect to solar abundance ratios }.
\label{demtablevel}
\begin{tabular}{lccccccccc}
\hline\hline
Model & $sc$ (MAX) & $sc$ (MIN) & $\chi^2_{red}$ (dof)\tablenotemark{a} & Ne & Mg & Si & S & Fe & $F_{\rm X}^{unabs}$ (tot, MIN--MAX) \\
& &  & & & & & & & ($10^{-13}$\,erg\,cm$^{-2}$\,s$^{-1}$) \\ 
\hline
no broadening  & 0.280$\pm$0.017 & 0.176$\pm$0.011 & 0.37(500) & 1.37$\pm$0.33 & 0.71$\pm$0.08 & 1.12$\pm$0.10 & 1.30$\pm$0.46 & 0.60$\pm$0.14 & 15.5--9.79\\
Gaussian broad.\tablenotemark{b}
               & 0.260$\pm$0.015 & 0.164$\pm$0.011 & 0.29(498) & 2.20$\pm$0.54 & 1.22$\pm$0.19 & 1.50$\pm$0.15 & 1.50$\pm$0.50 & 0.69$\pm$0.16 & 15.4--9.81\\
Pole-on model  & 0.275$\pm$0.017 & 0.173$\pm$0.011 & 0.33(500) & 1.50$\pm$0.36 & 0.84$\pm$0.10 & 1.24$\pm$0.11 & 1.37$\pm$0.48 & 0.63$\pm$0.14 & 15.4--9.71\\
Equ.-on model  & 0.279$\pm$0.017 & 0.175$\pm$0.012 & 0.33(500) & 1.56$\pm$0.37 & 0.85$\pm$0.10 & 1.26$\pm$0.12 & 1.37$\pm$0.48 & 0.64$\pm$0.15 & 15.4--9.69\\
\hline
\end{tabular}
\tablenotetext{a}{\ch\ data of maximum and minimum phases were fitted simultaneously, allowing for different scaling factors but forcing abundances to be the same: a single $\chi^2$ is thus provided for both phases. }
\tablenotetext{b}{For the gaussian broadening ({\sc gsmooth} model within XSPEC), the FWHMs were found to be 248$\pm$45\,\kms\ and 306$\pm$58\,\kms\ for maximum and minimum emission phases, respectively.}
\end{table*}

As a second step, we fitted the \ch\ high-resolution spectra, allowing for non-solar abundance in the elements whose lines are clearly seen in the HEG/MEG spectra (i.e., Ne, Mg, Si, S, Fe). Note that, for this exercise, the high-resolution spectra were binned in a similar way as the lower-resolution ones (see end of \S 2). Furthermore, to avoid the UV-depopulating effects modifying the $f/i$ ratios which are not considered in $apec$, the $f$ and $i$ lines of He-like triplets were grouped in a single bin. As the instrumental broadening of HEG/MEG spectra is much smaller than for low-resolution data, an intrinsic broadening can be more easily detected. Therefore, we tested several hypotheses: (1) no intrinsic broadening, (2) Gaussian broadening (whose amplitude was let free to vary), and (3) simulated line profiles (Fig. \ref{modelline}). The latter scenario allows us to perform a fully coherent comparison between data and 3D MHD simulations, as it uses the complete physical picture (simulated distribution of emissivity as a function of temperature {\it and} velocity) provided by the model. 

Results of these fits are provided in Table \ref{demtable} and shown for the best lines in Fig. \ref{demcompa2}. The scaling factors are similar to those found on lower-resolution spectra (indeed, a global fit to all \ch\ spectra also yields similar results). Derived abundances are quasi solar: indeed, the solar abundance is within 1--2$\sigma$ of the fitted value for Ne, Mg, Si, and S or within 3$\sigma$ for Fe. Besides, letting them freely vary only allows to (slightly) improve the $\chi^2$ (e.g. from 0.36 to 0.29 for the Gaussian broadening case). The fitting results thus show no clear and definitive evidence for non-solar abundances for these elements. A comparison between the different broadening hypotheses is more interesting. Even if the differences are marginal, note that the worst $\chi^2$ is obtained for no broadening, and the best one for Gaussian broadening. The FWHMs in this case are twice smaller than found on individual line analysis (but this remains within the errors, see Sect. 4.1) and, as in that analysis, the derived Gaussian broadening is again slightly larger at minimum emission phase, though the difference is marginal (within 2$\sigma$). The best-fit Gaussian broadening has a larger value than measured in the simulated profiles (see end of \S. 4.1 and Fig. \ref{modelline2}). This certainly indicates that the observed X-ray lines are broader than expected, as already derived from the line-by-line analysis. Yet, the very good agreement between observations and model predictions and the very limited improvement when considering a larger broadening are remarkable, showing that only further refinements of the model are still needed. 

\section{Conclusion}
We have obtained \ch\ data of the magnetic Of?p star \hd\ at two crucial phases (maximum and minimum emissions, when confined winds are seen face-on and edge-on, respectively). These new data show great similarities with \xmm-EPIC spectra (i.e. 40\% flux decrease and spectrum softening at minimum), demonstrating the quasi-perfect repeatability of the X-ray behavior over a decade. The high-resolution data further reveal more detail, with many similarities with HD\,148937 and $\theta^1$\,Ori\,C, the only two other magnetic O-stars observed at high-resolution: small (but non-zero) line broadenings for high-Z elements, negligible line shifts, hot plasma located at a few stellar radii from the star. In addition, comparing spectra of \hd\ at the two phases yields no significant change except for flux - the slightly larger broadening and slightly lower line shift found at minimum phase are only marginal, 1$\sigma$ changes, thus requiring confirmation with future X-ray facilities such as {\it Athena-XIFU}. 

We further compared the observational results with predictions from a dedicated 3D MHD simulation of confined winds in \hd. To this aim the simulated DEM was {\it directly} fitted to the observed spectra. The low-resolution data appear well fitted up to $\sim$3\,keV, a slight overprediction is seen at higher energies which can possibly be mitigated by including inverse Compton cooling in future models. At high-resolution, the X-rays lines also appear quite well fitted by the model, though a larger broadening yields slightly better results. A scaling of the total predicted flux by a factor of $\sim$5 is needed but this can be addressed by some reduction of the mass-loss rates, probably due to clumping of the wind. 

Refinements in the modeling are certainly needed, but the remarkable agreement between data and model certainly shows that the basic picture is promising. One avenue to investigate may be linked to asymmetries. Indeed, occultation of an axisymmetric equatorial structure located at the position of the X-ray emitting plasma cannot explain the observed flux variation of 40\%, while an asymmetric distribution, linked e.g. to a magnetic geometry more complicated than a simple centered dipole, may well do so. 

\acknowledgments
YN acknowledges support from the Fonds National de la Recherche Scientifique (Belgium), the Communaut\'e Fran\c caise de Belgique, the XMM PRODEX contract (Belspo), and an ARC grant for concerted research actions financed by the French community of Belgium (Wallonia-Brussels federation). AuD acknowledges support by NASA through Chandra Award numbers GO5-16005X, AR6-17002C and  G06-17007B issued by the Chandra X-ray Observatory Center which is operated by the Smithsonian Astrophysical Observatory for and behalf of NASA under contract NAS8-03060. ADS and CDS were used for preparing this document. 

%\facilities{CXO}

\end{document}